\documentclass[a4paper,twocolumn,aps,pra,showpacs]{revtex4}
\usepackage[latin9]{inputenc}
\usepackage{float}
\usepackage{amsmath}
\usepackage{amssymb}
\usepackage{graphicx}
\usepackage{esint}

\makeatletter

\providecommand{\tabularnewline}{\\}

\@ifundefined{textcolor}{}
{%
 \definecolor{BLACK}{gray}{0}
 \definecolor{WHITE}{gray}{1}
 \definecolor{RED}{rgb}{1,0,0}
 \definecolor{GREEN}{rgb}{0,1,0}
 \definecolor{BLUE}{rgb}{0,0,1}
 \definecolor{CYAN}{cmyk}{1,0,0,0}
 \definecolor{MAGENTA}{cmyk}{0,1,0,0}
 \definecolor{YELLOW}{cmyk}{0,0,1,0}
}

\usepackage{mathrsfs}\usepackage{subfigure}\usepackage{multirow}\usepackage[title,titletoc,header]{appendix}

\makeatother

\begin{document}

\title{Tuning a magnetic Feshbach resonance with spatially modulated laser
light}

\author{Yi-Cai Zhang$^{1,2}$, Wu-Ming Liu$^{1}$ and Hui Hu$^{2}$}

\affiliation{$^{1}$Beijing National Laboratory for Condensed Matter Physics,
Institute of Physics, Chinese Academy of Sciences, Beijing 100190,
China \\
 $^{2}$Centre for Quantum and Optical Science, Swinburne University
of Technology, Melbourne 3122, Australia}

\date{\today}
\begin{abstract}
We theoretically investigate the control of a magnetic Feshbach resonance
using a bound-to-bound molecular transition driven by spatially modulated
laser light. Due to the spatially periodic coupling between the ground
and excited molecular states, there exists a band structure
of bound states, which can uniquely be characterized by some extra
bumps in radio-frequency spectroscopy. With the increasing of coupling
strength, the series of bound states will cross zero energy and directly
result in a number of scattering resonances, whose position and width
can be conveniently tuned by the coupling strength of the laser light
and the applied magnetic field (i.e., the detuning of the ground molecular
state). In the presence of the modulated laser light, universal two-body bound states near zero-energy threshold still exist. However, compared with the case without modulation, the regime for such universal states is usually small. An unified formula which embodies the influence of the modulated coupling on the resonance width is given. The spatially modulated coupling also implies a local spatially
varying interaction between atoms. Our work proposes a practical way
of optically controlling interatomic interactions with high spatial
resolution and negligible atomic loss.
\end{abstract}


\maketitle

\section{Introduction }

Ultracold atoms provides an ideal platform to investigate and simulate
many-body problems of condensed-matter physics, e.g., the Mott insulator
transition \cite{Bloch}, magnetic phase transition \cite{Lewenstein},
because of their unprecedented controllability in purity and interatomic
interactions. There are a number of tools now available to tune the
interatomic interactions, such as magnetic and optical Feshbach resonances,
optical lattices, etc.

The magnetic Feshbach resonance - resulting from the hyperfine coupling
between two atomic states (i.e., open channel) and a molecular state
(closed channel) near zero energy - have been widely used \cite{Kohler,Chin},
allowing the realization of the long-sought crossover from a molecular
Bose-Einstein condensate (BEC) to a Bardeen-Cooper-Schrieffer (BCS)
superfluid and the investigation of interesting few-body physics such
as Efimov effects \cite{Braaten,Kraemer}.

The interaction between atoms can be also tuned by laser light near
photo-association transition, when two free atoms couple to an excited
molecular state \cite{Fedichev,Bohn,Bohn2}. This so-called optical
Feshbach resonance has been experimentally realized \cite{Fatemi,Theis}.
Compared with the magnetic Feshbach resonance, optical Feshbach resonance
could be used to control the interatomic interaction with high temporal
and spatial flexibility. In addition, the optical transition between
atomic states and molecular states is always available for most atomic
species. Hence, optical Feshbach resonance becomes crucial to control
the interatomic interaction for alkaline-earth atoms because of the
lack of magnetic structure in their ground states \cite{Ciurylo,Enomoto,Thalhammer}.
In a recent experiment of optical Feshbach resonance, optical standing
wave is used to couple atomic and molecular states of ytterbium-174
atoms, leading to a spatially modulated interaction between atoms
\cite{Yamazaki}. Theoretically, this spatially modulated interatomic
interaction was understood by using a two-channel model \cite{Qi}.
Future experiments on, e.g., the simulation of Hawking radiation in
cold atoms \cite{Garay,Carusotto}, the emission of solitons \cite{Rodas-Verde},
the dynamics of BEC collapse \cite{Dong,Yan}, the localized to delocalized
transition of solitons \cite{Abdullaev}, and the phase separation
of Bose and Fermi gases \cite{Chien}, all resulting from spatially
varying interactions, may benefit from the control of interatomic
interactions with high spatial and temporal resolutions.

However, due to the large light-induced atomic loss, the use of optical
Feshbach resonance is greatly limited. To reduce the loss, it has
been proposed to use alkaline-earth atoms with narrow inter-combination
line-width \cite{Blatt}. Alternatively, one may optically control
a magnetic Feshbach resonance by using a bound-to-bound transition
between two molecular states \cite{Bauer,Bauer2,Fu} or by using a
molecular dark state \cite{Wu}. Experimentally, the shift of the
magnetic Feshbach resonance position and the modification of the two-body
$s$-wave scattering length due to the bound-to-bound transition have
been demonstrated for both atomic Bose \cite{Bauer,Bauer2} and Fermi
gases \cite{Fu}, by using spatially uniform laser light.

In this work, we investigate the optical control of a magnetic Feshbach
resonance by using spatially \emph{varying} (i.e., standing-wave-like)
laser light, which drives the bound-to-bound transition between a
ground molecular state and an excited molecular state. This scheme
can directly be implemented in current experiments \cite{Bauer,Fu}
by replacing the uniform laser light with a standing-wave light. It
offers the ability to tune the interatomic interactions with a spatial
modulation at the sub-micron level. Compared with the previous spatial
modulation of interatomic interactions with optical Feshbach resonance
\cite{Yamazaki,Qi}, the major advantage of our scheme is that the
optical induced atomic loss would be significantly suppressed \cite{Bauer,Fu}.
As a result, our proposal provides a practical way to experimentally
realize spatially modulated interatomic interactions, for the purpose
of simulating related many-body problems. As we shall see, our scheme
also has the advantage of tuning the \emph{width} of Feshbach resonances,
with a great flexibility.

The rest of paper is organized as follows. In the next section (Sec.
II), we introduce the model Hamiltonian and calculate the energy bands
of bound states. The scattering states are also investigated and a
series of scattering resonances are obtained. In Sec. III, we present
a detailed analysis and discussion to our results. Sec. IV is devoted
to a summary of this work.

\section{Theoretical framework}

\subsection{Model Hamiltonian }

 In the absence of the bound-to-bound
molecular transition, the system can be described by the following
atom-molecule Hamiltonian \cite{Drummond,Timmermans,Timmermans2,Holland},

\begin{eqnarray}
 & H & =H_{0a}+H_{Ia}+H_{g}+H_{ag},\notag\\
 & H_{0a} & =\sum_{\sigma}\int d\vec{r}\psi_{\sigma}^{\dagger}(\vec{r})\left[-\frac{\hbar^{2}\nabla^{2}}{2m}-\mu\right]\psi_{\sigma}(\vec{r}),\notag\\
 & H_{Ia} & =U\int d\vec{r}\psi_{\uparrow}^{\dagger}(\vec{r})\psi_{\downarrow}^{\dagger}(\vec{r})\psi_{\downarrow}(\vec{r})\psi_{\uparrow}(\vec{r}),\notag\\
 & H_{g} & =\int d\vec{R}\phi^{\dagger}(\vec{R})\left[-\frac{\hbar^{2}\nabla^{2}}{2M}-2\mu+v_{g}\right]\phi(\vec{R}),\notag\\
 & H_{ag} & =\chi\int d\vec{R}\left[\phi^{\dagger}(\vec{R})\psi_{\uparrow}(\vec{R}/2)\psi_{\downarrow}(\vec{R}/2)+\textrm{H.c.}\right],
\end{eqnarray}
where $H_{0a}$ and $H_{Ia}$ are respectively the kinetic Hamiltonian
and interaction Hamiltonian of atoms with the field operator $\psi_{\sigma}(\vec{r})$
($\sigma=\uparrow,\downarrow$); $H_{g}$ is the Hamiltonian of molecules
in their ground state with the field operator $\phi(\vec{R})$ and
$v_{g}$ denotes the energy difference between the molecular state
and atomic state; $H_{ag}$ describes the atom-molecule coupling and
models the conversion between atoms and molecules. The mass of molecules
is twice of atomic mass $M=2m$. $\mu$ is the chemical potential.
$\textrm{H.c.}$ denotes the Hermitian conjugate. Note that we have
assumed short-range contact interactions for both interatomic interaction
$U(\vec{r}-\vec{r}')=U\delta(\vec{r}-\vec{r}')$ and atom-molecule
coupling $\chi(\vec{R};\vec{r},\vec{r}')=\chi\delta[\vec{R}-(\vec{r}+\vec{r}')/2]\delta(\vec{r}-\vec{r}')$.

We now consider the molecular bound-bound transition driven by a standing-wave
laser light $\Omega\cos(\vec{K}\cdot\vec{R})/2$, where $\Omega$
is the related Rabi frequency and $\vec{K}$ is the wave-vector of
the light. By using the field operator $\phi_{e}(\vec{R})$ for the
excited molecular state and taking the rotating-wave approximation,
we obtain the following two additional terms \cite{Bauer,Fu} :
\begin{eqnarray}
H_{e} & = & \int d\vec{R}\phi_{e}^{\dagger}(\vec{R})\left[-\frac{\hbar^{2}\nabla^{2}}{2M}-2\mu+v_{e}-\Delta-i\frac{\gamma}{2}\right]\phi_{e}(\vec{R}),\notag\\
H_{ge} & = & \int d\vec{R}\frac{\Omega\cos(KX)}{2}\left[\phi^{\dagger}(\vec{R})\phi_{e}(\vec{R})+\phi_{e}^{\dagger}(\vec{R})\phi(\vec{R})\right],
\end{eqnarray}
where $H_{e}$ is the kinetic Hamiltonian of the excited molecular
state, $v_{e}$ is the energy of the excited state relative to
the atomic state, $\Delta$ is the detuning of the molecular transition,
$\gamma$ describes the decay of the excited state, and $H_{ge}$
is the coupling between the ground and excited states through the
optical standing wave. We have assumed that the laser light is applied
along the $x$-direction so that $\cos(\vec{K}\cdot\vec{R})=\cos(KX)$.

In the case of large detuning ($\Delta\gg v_{e}$, $\gamma$), we
may safely neglect the decay of the excited molecular state (i.e.,
$\gamma=0$) and eliminate the field operator $\phi_{e}(\vec{R})$.
The coupling ($H_{ge}$) between molecular states leads to a Stark
energy shift $\Omega^{2}\cos^{2}(KX)/(4\Delta)$ for the molecular
ground state and consequently we have a modified Hamiltonian for ground-state
molecules,
\begin{align}
\tilde{H}_{g} & =\int d\vec{R}\phi^{\dagger}\left[-\frac{\hbar^{2}\nabla^{2}}{2M}-2\mu+v_{g}+\frac{\Omega^{2}\cos{}^{2}(KX)}{4\Delta}\right]\phi(\vec{R}).
\end{align}
It is obvious that the Stark energy shift plays the role of optical
lattices for ground-state molecules \cite{Fedichev2,Orso}. By taking
a Fourier transformation, the total Hamiltonian can be rewritten in
momentum space as
\begin{eqnarray}
H & = & H_{0a}+H_{Ia}+\tilde{H}_{g}+H_{ag},\notag\\
H_{0a} & = & \sum_{\vec{k}\sigma}(\epsilon_{\vec{k}}-\mu)C_{\vec{k}\sigma}^{\dagger}C_{\vec{k}\sigma},\notag\\
H_{Ia} & = & U\sum_{\vec{k},\vec{k}^{\prime},\vec{q}}C_{\vec{q}/2+\vec{k},\uparrow}^{\dagger}C_{\vec{q}/2-\vec{k},\downarrow}^{\dagger}C_{\vec{q}/2-\vec{k}^{\prime},\downarrow}C_{\vec{q}/2+\vec{k}^{\prime},\uparrow},\notag\\
\tilde{H}_{g} & = & \sum_{\vec{q}}\left(\frac{\epsilon_{\vec{q}}}{2}-2\mu+v_{g}+\frac{\Omega^{2}}{8\Delta}\right)b_{\vec{q}}^{+}b_{\vec{q}}\notag\\
 &  & -\sum_{\vec{q}}\frac{\Omega^{2}}{16\Delta}\left(b_{\vec{q}}^{\dagger}b_{\vec{q}+2K}+\textrm{H.c.}\right),\notag\\
H_{ag} & = & \chi\sum_{\vec{k},\vec{q}}\left(b_{\vec{q}}^{\dagger}C_{\vec{q}/2-\vec{k},\downarrow}C_{\vec{q}/2+\vec{k},\uparrow}+\textrm{H.c.}\right).
\end{eqnarray}
Here $\epsilon_{\vec{k}}=\vec{k}^{2}/2$ is the kinetic energy (in
the units of $m=1$ and $\hbar=1$). The above Hamiltonian will be
our starting point. In the following, we will solve the two-particle
problem of the Hamiltonian.

Note that, in the case of large detuning, the molecular excited state
$|e\rangle$ does not appear in the above Hamiltonian. Note also that,
here the lattice potential only appears for the ground molecular state,
unlike the case of an optical Feshbach resonance, where the spatial
modulation appears in the atom-molecule coupling $\chi$ \cite{Qi}.

\subsection{Two-body bound states }

Here we focus on the two-body problem, so the chemical potential $\mu=0$.
Due to the presence of the lattice potential, eigenstates can be classified
according to \emph{quasi-momentum} $q\in[-K,K]$
(Note that the period of the lattice in Eq. (3) is half of the wave
length of laser beam). Hereafter, $q$ and $K$ are understood as
along the $x$-direction unless explicitly specified. It is expected
that the eigen-energy would form a band structure. The two-body wave
function can be written as
\begin{align}
|\psi\rangle & =\sum_{n}A_{n}|nK+q,g\rangle\notag\\
 & +\sum_{n\vec{k}}B_{n,\vec{k}}|(nK+q)/2+\vec{k},\uparrow;(nK+q)/2-\vec{k},\downarrow\rangle,\label{eq:BoundState}
\end{align}
where $|nK+q,g\rangle$ is the molecular state with a center-of-mass
momentum $nK+q$, $|(nK+q)/2+\vec{k},\uparrow;(nK+q)/2-\vec{k},\downarrow\rangle$
is the state of a pair of atoms with total momentum $nK+q$ and relative
momentum $\vec{k}$ and with un-like spins. The two-particle Schr\"{o}dinger
equation reads
\begin{align}
 & H|\psi\rangle=E|\psi\rangle,
\end{align}
from which we determine coupled equations for the coefficients
$A_{n}$ and $B_{n,k}$,
\begin{eqnarray}
EA_{n} & = & \left[\epsilon_{nK+q,g}+v_{g}+\frac{\Omega^{2}}{8\Delta}\right]A_{n}\notag\\
 &  & -\frac{\Omega^{2}}{16\Delta}\left[A_{n+2}+A_{n-2}\right]+\chi\sum_{\vec{k}}B_{n,\vec{k}},\notag\\
EB_{n,\vec{k}} & = & \left[\epsilon_{(nK+q)/2-\vec{k},a}+\epsilon_{(nK+q)/2+\vec{k},a}\right]B_{n,\vec{k}}\notag\\
 &  & +U\sum_{\vec{k}^{\prime}}B_{n\vec{k}^{\prime}}+\chi A_{n},\label{eq:AnBnBoundState}
\end{eqnarray}
where the molecular kinetic energy $\epsilon_{nK+q,g}=(nK+q)^{2}/4$
and the atomic kinetic energy $\epsilon_{(nK+q)/2\pm\vec{k},a}=[(nK+q)/2\pm\vec{k}]^{2}/2$.
The above equation demonstrates that the molecular amplitudes of different
momenta $A_{n}$ are coupled by the lattice potential.
After eliminating the atomic amplitude $B_{n,\vec{k}}$, we obtain,
\begin{eqnarray}
EA_{n} & = & \left[\epsilon_{nK+q,g}+v_{g}+Z_{n}+\frac{\Omega^{2}}{8\Delta}\right]A_{n}\notag\\
 &  & -\frac{\Omega^{2}}{16\Delta}[A_{n+2}+A_{n-2}],
\end{eqnarray}
where,
\begin{align}
 & Z_{n}=\frac{\chi^{2}f_{n}}{1-Uf_{n}},\notag\\
 & f_{n}=\Sigma_{\vec{k}}\frac{1}{E-(\epsilon_{(nK+q)/2-\vec{k},a}+\epsilon_{(nK+q)/2+\vec{k},a})}.\notag
\end{align}
The bare parameters ($\chi$, $U$ and $v_{g}$) need to be renormalized
to real physical observables \cite{Fu}, for example,
\begin{align}
v_{g}+Z_{n}\rightarrow v_{g0}+Z_{n0}=v_{g0}+\frac{\chi_{0}^{2}f_{n0}}{1-U_{0}f_{n0}},
\end{align}
where
\begin{align}
f_{n0} & =\sum_{\vec{k}}\left[\frac{1}{E-[\epsilon_{(nK+q)/2-\vec{k},a}+\epsilon_{(nK+q)/2+\vec{k},a}]}+\frac{1}{\vec{k}^{2}}\right]\notag\\
 & =\sum_{\vec{k}}\left[\frac{1}{E-[(nK+q)^{2}/4+\vec{k}^{2}]}+\frac{1}{\vec{k}^{2}}\right]\notag\\
 & =\frac{\sqrt{-E+(nK+q)^{2}/4}}{4\pi}.\notag
\end{align}
Detailed expressions for real observables $v_{g0}$, $\chi_{0}$
and $U_{0}$ are given in the next section.

Eq. (8) differs from the usual eigenvalue problems in that the eigenvalue
$E$ appears on both sides of the equation. We can divide the eigenvalue
$E$ on both sides of the equation and obtain,
\begin{align}
A_{n}= & \frac{\left[\epsilon_{nK+q,g}+v_{g0}+Z_{n0}+\Omega^{2}/(8\Delta)\right]}{E}A_{n}\notag\\
 & -\frac{\Omega^{2}}{16\Delta E}\left[A_{n+2}+A_{n-2}\right].
\end{align}
The above equation has the form,
\begin{align}
 & |\psi\rangle=K(E)|\psi\rangle,
\end{align}
where the matrix elements of the kernel $K(E)$ depend on the eigenvalue
$E$. By adjusting $E$ to force the eigenvalues of the kernel $K(E)$
to be 1, we can solve all the eigenvalues and eigenvectors numerically.
Then, from the molecular amplitudes ($A_{n}$), one can obtain the
atomic amplitudes
\begin{align}
B_{n,\vec{k}}=\frac{\beta_{n}}{-E_{b}-[\epsilon_{(nK+q)/2-\vec{k},a}+\epsilon_{(nK+q)/2+\vec{k},a}]},\notag
\end{align}
where $E_{b}\equiv-E>0$ is the binding energy of the bound state
and
\begin{equation}
\beta_{n}\equiv\frac{U_{0}\chi_{0}f_{n0}A_{n}}{1-U_{0}f_{n0}}+\chi_{0}A_{n}.
\end{equation}

\subsection{Radio-frequency spectroscopy of two-particle bound states }

The existence of two-particle bound states may be detected by the
radio-frequency (rf) spectroscopy technique. The Hamiltonian of the
rf process can be written as \cite{Chin2,Bartenstein,Hu}
\begin{align}
V_{rf} & =V_{0}\int d\vec{r}\left[\psi_{3}^{\dagger}(\vec{r})\psi_{\downarrow}(\vec{r})+\textrm{H.c.}\right],\notag\\
 & =V_{0}\sum_{\vec{q}}\left[C_{\vec{q},3}^{\dagger}C_{\vec{q},\downarrow}+\textrm{H.c.}\right].
\end{align}
It represents a transition process, where the atoms in the state $|\vec{q},\downarrow\rangle$
are transferred to a third, unoccupied state $|\vec{q},3\rangle$.

Recall that the atomic part of the wave function of a two-particle
bound state is given by,
\begin{align}
|\psi,a\rangle & =\sum_{n\vec{k}}B_{n,\vec{k}}|(nK+q)/2+\vec{k},\uparrow;(nK+q)/2-\vec{k},\downarrow\rangle.
\end{align}
By acting $V_{rf}$ on this wave function, we obtain,
\begin{align}
 & V_{rf}|\psi,a\rangle\notag\\
 & =-V_{0}\sum_{n\vec{k}\vec{q}^{\prime}}B_{n\vec{k}}C_{\vec{q}^{\prime},3}^{\dagger}C_{(nK+q)/2+\vec{k},\uparrow}^{\dagger}C_{\vec{q}^{\prime},\downarrow}C_{(nK+q)/2-\vec{k},\downarrow}^{\dagger}|0\rangle,\notag\\
 & =-V_{0}\sum_{n\vec{k}}B_{n\vec{k}}C_{(nK+q)/2-\vec{k},3}^{\dagger}C_{(nK+q)/2+\vec{k},\uparrow}^{\dagger}|0\rangle,
\end{align}
which give us the final two-particle state after the rf pulse. Using
Fermi's Golden Rule, the transfer strength of the rf process is
given by the following Frank-Condon factor,
\begin{align}
\Gamma(\omega) & =\frac{1}{\mathcal{C}}\sum_{n,\vec{k}}|B_{n,\vec{k}}|^{2}\delta\left(\omega-\left[\frac{(nK+q)^{2}}{4}+\vec{k}^{2}+E_{b}\right]\right),
\end{align}
where the $\delta$-function guarantees energy conservation during
the rf process and $\mathcal{C}=\sum_{n,\vec{k}}|B_{n,\vec{k}}|^{2}$
is the normalization constant. By introducing $E_{n}=E_{b}+(nK+q)^{2}/4$,
we find that $f_{n0}=\sqrt{E_{b}+(nK+q)^{2}/4}/4\pi=\sqrt{E_{n}}/4\pi$,
$|B_{n\vec{k}}|^{2}=\beta_{n}^{2}/(E_{n}+\vec{k}^{2})^{2}$, and $\mathcal{C}=\sum_{n}\beta_{n}^{2}/[8\pi\sqrt{E_{n}}]$.
The Frank-Condon factor can then be rewritten as,
\begin{equation}
\Gamma(\omega)=\sum_{n}\frac{\beta_{n}^{2}}{4\pi^{2}\mathcal{C}}\frac{\sqrt{\omega-E_{n}}}{\omega^{2}}\theta(\omega-E_{n}),\label{eq:FC}
\end{equation}
where $\theta(x)$ is the Heaviside step function. Therefore, once
we obtain $E_{b}$ and $A_{n}$, the rf transfer strength can be calculated
straightforwardly.

\subsection{Two-particle scattering states}

We now consider the low-energy scattering state with energy $E>0$
and $E\ll K^{2}$. Here we focus on the isotropic $s$-wave scattering
at the quasi-momentum $q=0$. Without loss of generality, we assume
that the incident wave propagates along the $z$-direction. The scattering
wave function can be written as
\begin{eqnarray}
|\psi\rangle & = & |k_{z},\uparrow;-k_{z},\downarrow\rangle+\sum_{n}A_{n}|nK,g\rangle+\notag\\
 &  & +\sum_{n,\vec{k}}B_{n,\vec{k}}|nK/2+\vec{k},\uparrow;nK/2-\vec{k},\downarrow\rangle,
\end{eqnarray}
where the first term on the right-hand side $|k_{z},\uparrow;-k_{z},\downarrow\rangle$
stands for the incident state of two atoms with the total momentum
$0$, relative momentum $k_{z}$ and energy $E=k_{z}^{2}$. By substituting
the wave function into the two-particle Schr\"{o}dinger equation, we obtain,
\begin{eqnarray}
EA_{n} & = & \left[\epsilon_{nK,g}+v_{g}+\frac{\Omega^{2}}{8\Delta}\right]A_{n}-\frac{\Omega^{2}}{16\Delta}\left[A_{n+2}+A_{n-2}\right]\notag\\
 &  & +\chi\sum_{\vec{k}}B_{m,\vec{k}},\notag\\
EB_{n,\vec{k}} & = & \left[\epsilon_{nK/2-\vec{k},a}+\epsilon_{nK/2+\vec{k},a}\right]B_{n,\vec{k}}\notag\\
 &  & +U\sum_{\vec{k}^{\prime}}B_{n\vec{k}^{\prime}}+\chi A_{n}+U\delta_{n,0}.
\end{eqnarray}
Here, compared with Eq. (\ref{eq:AnBnBoundState}), the extra term
$U\delta_{n,0}$ in the last line comes from the incident state with
zero total momentum.

It is important to note that, traditionally, in the absence of optical
lattices the atomic and molecular states are referred to as the open
and closed channels, respectively. In our case with the lattice potential,
this two-channel viewpoint should be generalized, as the dispersion
relation is now folded into discrete energy bands (i.e., different
$n$). That is, we may classify any atomic states with a nonzero band
index $n\neq0$ as a closed channel \cite{Qi}. As a result, with the lattice
potential we are now dealing with a \emph{multi-channel} scattering
problem, instead of the usual two-channel problem. As we shall see
later, this multi-channel viewpoint is crucial to understand the width
of scattering resonances.

By adopting the similar strategy of eliminating the atomic amplitudes
$B_{n,\vec{k}}$ as in the bound state calculation, we obtain,
\begin{eqnarray}
EA_{n} & = & \left[\epsilon_{nK,g}+v_{g}+Z_{n}+\frac{\Omega^{2}}{8\Delta}\right]A_{n}\notag\\
 &  & -\frac{\Omega^{2}}{16\Delta}\left[A_{n+2}+A_{n-2}\right]+\frac{Z_{n}U}{\chi}\delta_{n,0},
\end{eqnarray}
After the renormalization, the equation becomes
\begin{eqnarray}
EA_{n}& = \left[\epsilon_{nK,g}+v_{g0}+Z_{n0}+\frac{\Omega^{2}}{8\Delta}\right]A_{n}\notag\\
   & -\frac{\Omega^{2}}{16\Delta}\left[A_{n+2}+A_{n-2}\right]+\left(\frac{Z_{n0}U_{0}}{\chi_{0}}+\chi_{0}\right)\delta_{n,0}.
\end{eqnarray}
We can solve the above linear equation to obtain the molecular amplitudes
$A_{n}$, and then the atomic amplitudes $B_{nk}$ through the expression
$B_{nk}=\beta_{n}^{\prime}/[-(E_{n}^{\prime}+\vec{k}^{2})]$, where
$E_{n}^{\prime}=-E+(nK)^{2}/4$ and
\begin{equation}
\beta_{n}^{\prime}=\frac{U_{0}^{2}f_{n0}\delta_{n,0}+U_{0}x_{0}f_{n0}A_{n}}{1-U_{0}f_{n0}}+x_{0}A_{n}+U_{0}\delta_{n,0}.
\end{equation}

\subsection{Spatially modulated interatomic interactions}

In coordinate space, the atomic part of the scattering wave function
can be written as,
\begin{align}
\langle r|\psi,a\rangle & =e^{ik_{z}z}+\sum_{n,\vec{k}}B_{n,\vec{k}}e^{inKX+i\vec{k}\vec{r}},\notag\\
& =e^{ik_{z}z}+\sum_{n}\frac{1}{(2\pi)^3}\int d^3\vec{k} \frac{-\beta^{\prime}_{n}}{E^{\prime}_n+\vec{k}^2}e^{inKX+i\vec{k}\vec{r}},\notag\\
 & =e^{ik_{z}z}-\frac{\beta_{0}^{\prime}e^{i\sqrt{E}r}}{4\pi r}+\sum_{n\neq0}-\frac{\beta_{n}^{\prime}e^{inKX}e^{-\sqrt{E_{n}^{\prime}}r}}{4\pi r}.\label{eq:wfs1}
\end{align}
We can see that the wave functions of closed channels ($n\neq0$)
all exponentially decrease with increasing $r$. As the incident energy
$E\rightarrow0$, the $s$-wave scattering amplitude is given by $f_{0}=-\beta_{0}^{\prime}/4\pi$,
from which we determine the $s$-wave scattering length
\begin{equation}
a_{eff}=-f_{0}=\frac{\beta_{0}^{\prime}}{4\pi}.
\end{equation}
On the other hand, at the short range ($r\rightarrow0$), the atomic
part of the scattering wave function can be expressed as
\begin{align}
\langle r|\psi,a\rangle & \propto1/r-1/a_{loc}(X)+o(r),\label{eq:wfs2}
\end{align}
where $o(r)$ represents a quantity at the same order of magnitude
of $r$, and $a_{loc}(X)$ can be interpreted as the \emph{local}
$s$-wave scattering length. Comparing Eq. (\ref{eq:wfs1}) with Eq.
(\ref{eq:wfs2}), we obtain the expression of the local $s$-wave
scattering length \cite{Qi}
\begin{align}
a_{loc}(X) & =\frac{1-\sum_{n\neq0}U_{n}\cos(nKX)/U_{0}}{1/a_{eff}-\sum_{n\neq0}U_{n}|n|K\cos(nKX)/U_{0}},
\end{align}
where
\begin{eqnarray}
U_{0} & = & +\frac{\beta_{0}^{\prime}}{4\pi},\notag\\
U_{n} & = & -\frac{\beta_{n}^{\prime}}{4\pi}.
\end{eqnarray}
Note that, when we construct an effective many-body Hamiltonian of
our system, the interaction Hamiltonian may be modeled by using the
local scattering length $a_{loc}(X)$ \cite{Qi}, which is position
dependent. Thereby, the lattice potential gives rise to a spatially
modulated interatomic interaction.

\section{Results and discussion}

Taking an ultracold Fermi gas of $^{40}$K atoms as an example \cite{Fu},
at the magnetic Feshbach resonance $B_{0}=202.20\pm0.02$ G the background
scattering length $a_{bg}\simeq174a_{B}$ ($a_{B}$ is the Bohr radius),
the difference in magnetic momentum of atoms and of ground-state molecules
is $\mu_{ag}=2\mu_{a}-\mu_{g}\simeq2\mu_{B}$ ($\mu_{B}$ is the Bohr
magneton), and the width of resonance $W\simeq7.04\pm0.10$ G. In
the following calculations, we take the natural units: the mass of
atoms $m=1$, the background scattering length $a_{bg}=1$ and $\hbar=1$.
Therefore, energy is measured in units of $\hbar^{2}/ma_{bg}^{2}$.
We take the parameters: $B-B_{0}=-0.6$ G; $\Omega=2\pi\hbar\times0.07$
GHz; the wave length of laser $\lambda=780$ nm; the wave vector $K=2\pi/\lambda$.
The physical observables mentioned earlier are related to the above
experimental parameters by the expressions,
\begin{eqnarray}
U_{0} & = & 4\pi\hbar^{2}a_{bg}/m,\notag\\
\chi_{0} & = & 2\hbar\sqrt{\pi a_{bg}W\mu_{ag}/m},\notag\\
\nu_{g0} & = & \mu_{ag}(B-B_{0}).
\end{eqnarray}

\begin{figure}[H]
\centering \includegraphics[scale=0.5]{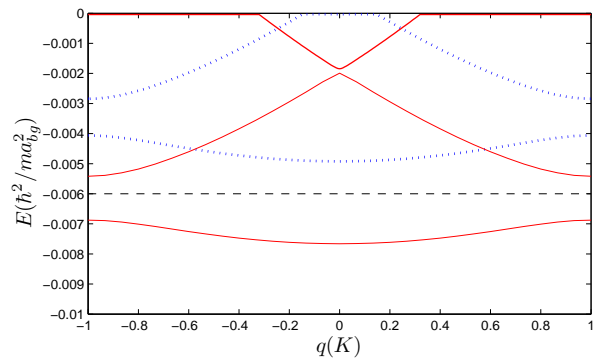}\\[0pt] \caption{(Color online). The bound states energy spectrum with optical coupling
$\Omega^{2}/16\Delta=0.0342$ (dotted blue lines) and $-0.0342$ (solid red lines),
respectively. The dashed line indicates the lowest energy of bound
states without lattice potential (i.e., $\Omega^{2}/16\Delta=0$). }

\label{fig1}
\end{figure}

\begin{figure}[H]
\centering \includegraphics[scale=0.5]{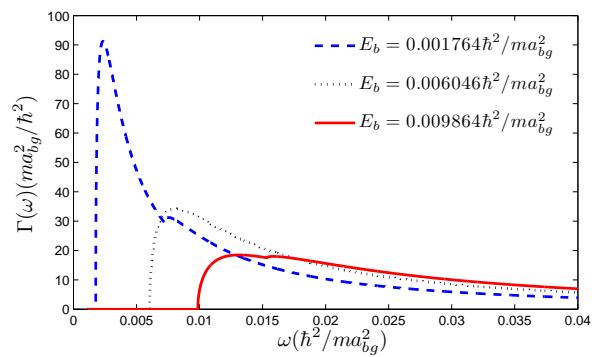}\\[0pt] \caption{(Color online). The Frank-Condon factor. The dashed blue, dotted black and solid red lines in Fig. \ref{fig2} correspond bound states 1, 2 and 3 in Fig. \ref{fig3} (a), respectively {[}see the three bound states denoted
by arrow heads on the rightmost line of the panel (a) of Fig. \ref{fig3}{]}.  }

\label{fig2}
\end{figure}

In Fig. \ref{fig1} we show the bound state energies. The dotted blue and
solid red lines correspond to the blue ($\Delta>0$) and red detunings ($\Delta<0$),
respectively. As anticipated, overall the red detuning gives rise
to a lower energy for two-particle bound states. Fig. \ref{fig2}
reports the rf spectroscopy of three lowest bound states located
at quasi-momentum $q=0$ for different lattice depths.
We find that the smaller binding energy is, the sharper is rf signal (see
the dashed blue line in Fig. \ref{fig2}). This is because when the binding energy approach zero, the wave function of bound states extends widely in coordinate space. Accordingly, the wave function in momentum space will concentrate near zero-momentum. So the overlap of wave functions which gives the Frank-Condon factor reaches large value near zero-energy.
 Due to the coupling of different
total momenta, for each bound state its atomic part of the wave-function
is a linear superposition of different components with different total
momenta, as shown in Eq. (\ref{eq:BoundState}). This results in additional
bumps in the rf spectroscopy, see for example, Eq. (\ref{eq:FC}).
Therefore, we may identify the bumps as a unique characteristic of
the energy band structure due to the lattice potential. With increasing
the lattice potential strength, the bumps becomes more evident. The
result of Fig. \ref{fig2} can be directly verified in current cold-atom
experiments by using rf-spectroscopy \cite{Bartenstein,Fu}.

\begin{figure}[H]
\centering \includegraphics[scale=0.5]{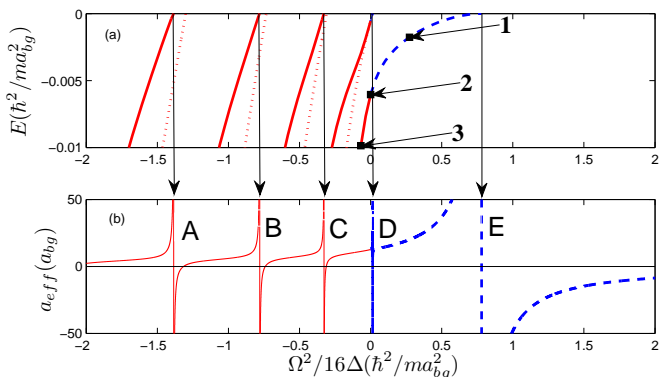}\\[0pt] \caption{(Color online). The bound states and their corresponding Feshbach
resonances at the quasi-momentum $q=0$. The panel (a) gives the evolution
of bound state energies with increasing the lattice depth. The panel
(b) shows the $s$-wave scattering length ($a_{eff}$). The solid red (dashed blue)
lines correspond to the case of $\Delta<0$ ($\Delta>0$). In (a),
the dotted lines show the energy branches that do not induce resonance
when they cross zero-energy. }

\label{fig3}
\end{figure}

To show the evolution of the energy band structure as a function of
the lattice depth $\Omega^{2}/16\Delta$, we report $E(q=0)$ in Fig.
\ref{fig3}. With increasing the lattice depth, in the case of red
detuning ($\Delta<0$), more bound states emerge {[}see the panel
(a) of Fig. \ref{fig3}{]}, while in the case of blue detuning ($\Delta>0$),
the energy of bound states move upward and crosses zero-energy. The
corresponding evolution of the $s$-wave scattering length is shown
in the panel (b) of Fig. \ref{fig3}.  A resonance occurs
in the $s$-wave scattering length when the energy of bound state
crosses zero-energy, as one may anticipate. However, not all the energy
branches induce resonance when they cross zero-energy. In the panel
(a), the wave function of the bound states shown in dotted red lines
is antisymmetric with respect to the momentum $nK=0$ (i.e., $\beta_{n}^{\prime}=-\beta_{-n}^{\prime}$),
implying $\beta_{0}^{\prime}=0$. As a result, the $s$-wave scattering
length $a_{eff}=\beta_{0}^{\prime}/4\pi=0$ and hence these bound
states do not result in any Feshbach resonance. We note that, the
resonance induced by a spatially modulated atom-molecule coupling
has been previously discussed in the case of optical Feshbach resonance
\cite{Qi}. The appearance of a resonance was similarly found to depend
on the symmetry of the bound state.

\begin{table}
\begin{tabular}{|c|c|c|c|c|c|c|}
\hline
 No. & A & B & C & D & E\tabularnewline
\hline
$(\Omega^{2}/16\Delta)_{0}$  & $-1.385$  & $-0.78$  & $-0.3279$  & $0.01399$  & $0.7807$ \tabularnewline
\hline
$W$  & $0.3526$  & $0.2535$  & $0.1403$  & $0.0045$  & $10.4948$ \tabularnewline
\hline
$E_{\text{uni}}$  & $5.1*10^{-5}$  & $3.3*10^{-5}$  & $1.1*10^{-5}$  & $7.6*10^{-9}$  & $0.0046$ \tabularnewline
\hline
\end{tabular}

\caption{The resonance position and width of the five Feshbach resonances shown
in the panel (b) of Fig. \ref{fig3}(in units of $\hbar^{2}/ma_{bg}^{2}$). $E_{\text{uni}}$ is an energy scale associated with the regime for universal two-body bound states (see Eq. (A.13) in the Appendex)}
\end{table}

\begin{figure}[H]
\centering \includegraphics[scale=0.5]{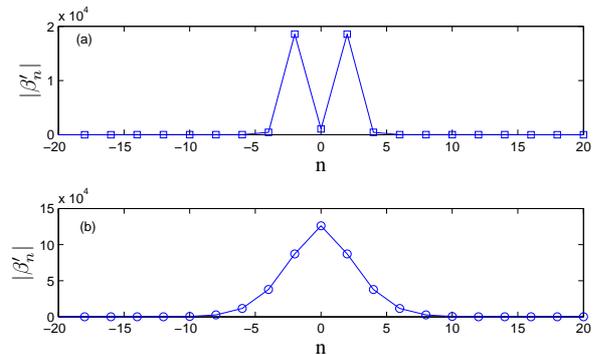}\\[0pt] \caption{(Color online). The atomic amplitudes as a function of the band index
$n$ near the Feshbach resonances \textbf{D} and \textbf{E} (corresponding
to the two resonances with blue detuning in the panel (b) of Fig.
\ref{fig3}). From the upper panel (a) for the resonance \textbf{D},
we see that the closed channels ($n=\pm2$) have the largest amplitudes.
Therefore, we interpret the resonance \textbf{D} as a closed-channel-dominated
resonance. On the contrary, the lower panel (b) for the resonance
\textbf{E} corresponds to a entrance-channel-dominated resonance.}

\label{fig4}
\end{figure}

Different from the case of optical Feshbach resonance \cite{Qi},
however, in our case the width of the spatial-modulation-induced resonance
varies significantly by changing the depth of the optical lattice
potential. Near resonance, the scattering length can be written as
\begin{equation}
a_{eff}=a_{bg}\left[1-\frac{W}{(\Omega^{2}/16\Delta)-(\Omega^{2}/16\Delta)_{0}}\right],
\end{equation}
here {[}$(\Omega^{2}/16\Delta)_{0}${]}
and ($W$) are the resonance position and width. In Table I, we calculate the width
of the five Feshbach resonances shown in the panel (b) of Fig. \ref{fig3} (for details see Appendix). The position of resonance can be obtained through fitting our numerical data.
 Generally, the widths of Feshbach resonance are influenced greatly by the other atomic closed channels (see Table. I). In the absence of optical lattice potential, the resonance width of
$^{40}$K atoms near the magnetic field $B_{0}=202.20\pm0.02$ is
about $W\sim3.3$, in the energy unit of $\hbar^{2}/ma_{bg}^{2}$. From
Table. I, we find that, in the presence of the lattice potential,
the resonance width can be one order of magnitude larger or smaller
than that without the lattice potential. For large blue detuning, the width of resonance \textbf{E} is extremely large. For red detuning
we find that the width becomes larger with increasing the depth of
the optical coupling $|\Omega^{2}/\Delta|$. As a result, we can access
very wide Feshbach resonance by choosing the zero-energy bound state
at large lattice depth.

Fig. \ref{fig4} reports the atomic amplitude $\left|\beta_{n}^{\prime}\right|$
near the Feshbach resonance \textbf{D} and \textbf{E} {[}see the panel
(b) of Fig. \ref{fig3}{]}. The width of the resonance \textbf{D}
is very small. It is a closed-channel-dominated resonance, in the
sense that the atomic amplitudes of closed channels $\beta_{n=\pm2}^{\prime}$
take the largest value relative to the open channel ($\beta_{n=0}^{\prime}$)
{[}see Fig. \ref{fig4}(a){]}. On the contrary, the resonance \textbf{E}
has a very large resonance width and the atomic amplitude peaks at
$n=0$.

\begin{figure}[H]
\centering \includegraphics[scale=0.5]{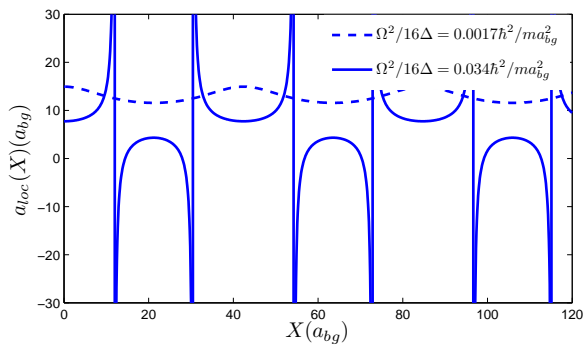}\\[0pt] \caption{(Color online). The local $s$-wave scattering length near the resonance
\textbf{D}. The solid and dashed lines have the scattering length
$a_{eff}=12.88a_{bg}$ and $a_{eff}=12.81a_{bg}$, respectively.}

\label{fig5}
\end{figure}


It is worth noting that although there is a modulated lattice, universal two-body bound states near
zero-energy threshold still exist (see Appendix), whose energy is approximately $E\propto -1/a^2_{eff}$. However, the universal regime may be extremely small because of the influence of other atomic closed channels. In Table. I, we calculate an characteristic energy scale $E_{\text{uni}}$ for each resonance, which determines the size of the universal regime. Only when the energy satisfy $|E|\ll E_{\text{uni}}$, the universal expression $E\propto -1/a^2_{eff}$ is valid (see Appendix). From Table. I, we find that the universal regimes for Feshbach resonances \textbf{A}, \textbf{B}, \textbf{C} and \textbf{D} are all extremely small. This explains why we can not see the universal behavior from Fig. 3. However, the Feshbach resonance \textbf{E} has a relatively large universal regime compared with others. As a result, the corresponding energy curve looks like quadratic parabola near the Feshbach resonance \textbf{E}.

Fig. 5 shows the spatial dependence of the local $s$-wave scattering
length. It is easy to see that the variation period of the scattering
length $a_{loc}(X)$ is directly determined by the optical lattice
potential. For a weak lattice potential (dashed line), the variation
of the local scattering length follows a cosine function. The mean
value of the local scattering length is roughly equal to the $s$-wave
scattering length $a_{eff}$. For a stronger lattice potential (solid
line), although $a_{eff}$ is nearly the same, the value of the local
scattering length changes drastically, from positive to negative,
when the position $X$ changes. This implies that a reasonably large
lattice potential has crucial effects on spatially modulated interatomic
interactions, similar to what has already been seen in
the case of an optical Feshbach resonance \cite{Qi}

\section{summary}

In conclusion, we have investigated how to tune a magnetic Feshbach
resonance by using standing-wave laser light that drives a molecular
bound-to-bound transition. The two-particle bound states and scattering
states (or scattering lengths) are significantly affected by the standing-wave
light. A band structure is formed and a series of zero-energy scattering
states appear. As a result, a number of laser-induced Feshbach resonances
emerge, whose position and width can be tuned by changing the depth
of the standing-wave laser. The resulting $s$-wave scattering length
near resonance shows a strong spatial dependence. This provides a
new tool to control interatomic interactions and therefore opens a
new route to study many interesting many-body physics, for example,
the exotic soliton, spatially inhomogeneous BCS superfluidity or BEC-BCS
crossover, self-trapping of BECs induced by spatially modulated interatomic
interactions.

Our proposed scheme can be directly examined in current experiments
for an ultracold Fermi gas of $^{40}$K atoms. Indeed, the optical
control of the interaction between $^{40}$K atoms near the broad
Feshbach resonance $B_{0}=202.20\pm0.02$ has recently been demonstrated
\cite{Fu}, by using a spatially homogeneous laser. Our scheme is
straightforward to implement by replacing the homogeneous laser with
a standing-wave laser. The predicted energy band structure and the
series of laser-induced Feshbach resonances could be easily observed
by using the radio-frequency spectroscopy and atomic loss spectroscopy.
We note that our calculations apply to bosonic systems as well. In
that case, the spatially modulated interatomic interaction can be
observed through the measurement of the mean-field energy of BECs
\cite{Yamazaki}.

\begin{acknowledgments}
This work was supported by the NKBRSFC under grants Nos. 2011CB921502, 2012CB821305, NSFC under grants Nos. 61227902, 61378017, 11434015, SKLQOQOD under grants No. KF201403, SPRPCAS under grants No. XDB01020300.
H.H. was supported by the Australian Research Council (ARC) Discovery Projects (Grant Nos. FT130100815 and DP140103231).
 \end{acknowledgments}

\appendix*

\section{The universal two-body bound states near zero-energy}
In this appendix, we show the existence of universal two-body bound states near zero-energy and discuss the size of the universal regime. At the same time, an explicit formulation for the resonance width is given.

We start by rewriting Eq.(10) in the form of an eigen equation:
\begin{align}
H(E(\lambda),\lambda)|\psi(\lambda)\rangle=E(\lambda)|\psi(\lambda)\rangle,
\end{align}
where $\lambda\equiv\Omega^2/16\Delta$ denotes the strength of the modulated lattice.
The $\lambda$-dependence of the Hamiltonian, wave function and energy has been explicitly emphasized.
The non-zero matrix elements of the Hamiltonian are
\begin{align}
&(H)_{n, n}=\epsilon_{nK+q,g}+v_{g0}+Z_{n0}+2\lambda,\notag\\
&(H)_{n, n+2}=(H)_{n+2, n}=-\lambda.
\end{align}
Let us take the derivative of Eq. (A.1) with respect to $\lambda$,
\begin{align}
\frac{\partial H}{\partial E}\frac{\partial E}{\partial \lambda}|\psi\rangle+\frac{\partial H}{\partial \lambda}|\psi\rangle+H|\frac{\partial \psi}{\partial \lambda}\rangle=\frac{\partial E}{\partial\lambda}|\psi\rangle+E|\frac{\partial \psi}{\partial \lambda}\rangle.
\end{align}
By acting $\langle \psi|$ on both sides of the above equation, we obtain,
\begin{align}
\frac{\partial E}{\partial\lambda}=\frac{\langle\psi|\frac{\partial H}{\partial\lambda}|\psi\rangle}{1-\langle\psi|\frac{\partial H}{\partial E}|\psi\rangle}.
\end{align}
It is easy to see that the non-zero matrix elements of $\frac{\partial H}{\partial \lambda}$ are
\begin{align}
&(\frac{\partial H}{\partial \lambda})_{n, n}=2,\notag\\
&(\frac{\partial H}{\partial \lambda})_{n, n+2}=(H)_{n+2, n}=-1.
\end{align}
Similarly, we have the non-zero matrix element of $\frac{\partial H}{\partial E}$,
\begin{align}
(\frac{\partial H}{\partial E})_{n, n}&=\frac{\partial Z_{n0}}{\partial E}=\frac{-\chi^2_0}{32\pi^2}\frac{1}{f_{n0}(1-U_0f_{n0})^2}.
\end{align}
The denominator in Eq. (A.4) is
\begin{align}
1-\langle\psi|\frac{\partial H}{\partial E}|\psi\rangle=1-\sum_n |A_n|^2\frac{\partial Z_{n0}}{\partial E}\equiv x_1+x_2,
\end{align}
here we have introduced $x_1\equiv1-\sum_{n\neq0} |A_n|^2\frac{\partial Z_{n0}}{\partial E}$ and $x_2\equiv-|A_0|^2\frac{\partial Z_{00}}{\partial E}$. Using the normalization of wave function ($\sum_n |A_n|^2=1$), the numerator is
\begin{align}
\langle\psi|\frac{\partial H}{\partial \lambda}|\psi\rangle&=2\sum_n |A_n|^2-\sum_n(A_nA_{n+2}+A_nA_{n-2})\notag\\
&=2-\sum_n(A_nA_{n+2}+A_nA_{n-2})\equiv C(\lambda).
\end{align}

Focusing on the case with $q=0$, when $E\rightarrow 0_-$, we know that $\frac{\partial Z_{00}}{\partial E}\approx-\frac{\chi^2_0}{8\pi}\frac{1}{\sqrt{-E}}\propto1/\sqrt{-E}$ diverges, while $\frac{\partial Z_{n0}}{\partial E}$ is finite for $n\neq0$ (see Eq. (A.6) and Eq. (9)). Thus, as long as $A_0\neq0$, the denominator is dominated by $x_2$, as the energy approaches zero, and Eq.(A.4) becomes
\begin{align}
\frac{\partial E}{\partial \lambda}=\frac{C(\lambda)}{x_1+x_2}\approx\frac{C(0)}{x_2}\approx\frac{C(0)}{\chi^2_0|A_0|^2/8\pi}\sqrt{-E}.
\end{align}
Here we assume the dependence on $\lambda$ in $C(\lambda)$ is weak and replace $C(\lambda)$ by  $C(0)\equiv \text{lim}_{E\rightarrow0}C(\lambda)$.
From the above equation, we obtain,
\begin{align}
E=-d^2(\lambda-\lambda_0)^2= -1/a_{eff}^2,
\end{align}
where $d\equiv\frac{4\pi C(0)}{\chi^2_0|A_0|^2}$. Compared with Eq.(29), we find that the resonance width is given by,
\begin{align}
W=1/d=\frac{\chi^2_0}{8\pi }\frac{2|A_0|^2}{C(0)}.
\end{align}

Eq.(A.10) demonstrates that, even in the presence of the modulated lattice, universal two-body bound states near zero-energy still exist. In the absence of the modulated lattice, the non-zero wave amplitude is $A_0=1$, so $C(0)=2$. Using Eq. (A.11), the resonance width is reduced to the two-channel limit $W=\chi^2_0/8\pi\approx 3.3 $, in units of $\hbar^2/ma^2_{bg}$. Thus, the factor $2|A_0|^2/C(0)$ embodies the influence of atomic closed channels on the width.
We have calculated the resonance widths near the five Feshbach resonances, as shown in Table I. For the resonance \textbf{E}, the wave-function amplitude $A_0$ is large and the other $A_n$ has the same sign as $A_0$. As a result, $C(0)$ is small (see Eq. (A.8)). The large factor $2|A_0|^2/C(0)$ results in a relatively large resonance width. For other resonances, due to the small $A_0$, the small factor $2|A_0|^2/C(0)$ gives a small resonance width.

The universal regime may be extremely narrow compared with the two-channel case.
From Eq. (A.9), the universal regime is given by the condition
\begin{align}
|x_1|\ll|x_2|=\frac{\chi^2_0|A_0|^2}{8\pi\sqrt{-E} },
\end{align}
so we have
\begin{align}
|E|\ll E_{\text{uni}},
\end{align}
where $E_{\text{uni}}\equiv(\frac{\chi^2_0|A_0|^2}{8\pi|x_1|})^2$.
In Table I, we list the energy scale $E_{\text{uni}}$ near the five zero-energy Feshbach resonances. We find that the universal regime is extremely small except for the resonance E. In the two-channel case without the modulated lattice ($|A_0|=1, |x_1|=1$), the energy scale $E_{\text{uni}}=(\frac{\chi^2_0}{8\pi})^2=10.89$, which is much larger than the energy scale for the five resonances (see the bottom line in Table I). In this sense, the universal regime of Feshbach resonances in the presence of the modulated lattice is always very small.


\end{document}